\documentclass[12pt,aps]{article}
\usepackage{amssymb}
\usepackage{amsmath}
\newcommand{\ket}[1]{{|#1\rangle}}
\newcommand{\bra}[1]{{\langle#1|}}
\newcommand\Lu{{\em L\"uders}}
\newcommand{\bc}[2]{\left(\begin{array}{cc}
    #1 \\ #2 \end{array} \right)}
\newcommand\tr{\mbox{ Tr }}
\newcommand\clebsch[6]{\left<\begin{array}{cc|c}
   #1 & #2 & #3 \\ #4 & #5 & #6 \end{array} \right>}

\newcommand\be{\begin{equation}}
\newcommand\ee{\end{equation}}
\newcommand\bea{\begin{eqnarray}}
\newcommand\eea{\end{eqnarray}}

\begin{document}
\noindent{\Large \sf L\"{u}ders Theorem for Coherent-State POVMs}\\ \\

\noindent{\sc \normalsize Stefan Weigert {\rm and} Paul Busch} \\ \\
\noindent{\normalsize \rm $\bra{\mbox{Hu}}\mbox{MP}\rangle$ -
\rm Department of Mathematics, University of Hull\\
UK-Hull HU6 7RX, United Kingdom \\
{\tt S.Weigert@hull.ac.uk}, {\tt P.Busch@hull.ac.uk}\\
August 2003 \\ \\

\parbox[height]{13cm}{\small {\bf Abstract}:
\Lu ' theorem states that two observables commute if measuring one
of them does not disturb the measurement outcomes of the other. We
study measurements which are described by continuous positive
operator-valued measurements (or POVMs) associated with coherent
states on a Lie group. In general, operators turn out to be
invariant under the \Lu\ map if their $P$- and $Q$-symbols
coincide. For a spin corresponding to $SU(2)$, the identity is
shown to be the only operator with this property. For a particle,
a countable family of linearly independent operators is identified
which are invariant under the \Lu\ map generated by the coherent
states of the Heisenberg-Weyl group, $H_3$. The \Lu\ map is also
shown to implement the anti-normal ordering of creation and
annihilation operators of a particle.}

\section*{\sf 1. Introduction}
In this paper we determine operators $B$ which are invariant under
a generalized {\em L\"uders} map
\be
B \mapsto \Lambda (B) = \int_{\mathbb{X}} d \mu(\Omega) \,
                          E(\Omega) B E(\Omega) \, ,
\label{general} \ee
where the set $E(\Omega)$ is a family of projection operators
 labelled by the points $\Omega$ of a
manifold $\mathbb{X}$. These operators constitute a continuous
positive operator-valued measure, or POVM, with a resolution of
unity:
\be
\int_{\mathbb{X}} d \mu(\Omega) \, E(\Omega) = I \, .
\label{contcomp} \ee
Any operator $B$, bounded or not, will be called \Lu\  if it is
invariant under {\em L\"uders}' map,
\be
\Lambda (B)= B \, .
\label{inv} \ee
The operator $B$ acts on a complex separable Hilbert space ${\cal
H}$, and the operator $E(\Omega)$ is a member of a (over-)
complete family of projectors on coherent states $\ket{\Omega}$
associated with an irreducible, unitary representation of a Lie
group $G$ in the space ${\cal H}$.

This setting generalizes the traditional approach to minimally
disturbing (or {\em ideal}) \Lu\ measurements. Given a
self-adjoint operator with spectral decomposition $A=\sum_i^N a_i
E_i$, $N\leq \infty$, the projectors $E_i$ are complete and
orthogonal,
\be
\sum_{i=1}^N E_i = I \, , \qquad E_i E_j = E_i \delta_{ij} \, ,
\quad i,j = 1, \ldots, N \leq \infty \, . \label{complete0} \ee
If a non-selective, ideal measurement of $A$ is performed on a
quantum system with density operator $\rho$, its state undergoes a
\Lu\ transformation:
\be
\rho \mapsto \Lambda (\rho) = \sum_{i=1}^N E_i \rho E_i \, ,
\label{Lutrf} \ee
which extends to a linear, completely positive map. If, for some
operator $B$, one has
\be
\tr{\left[\rho B\right]} = \tr{\left[\Lambda(\rho) B\right]} \, ,
\quad \mbox{ for all } \rho \, ,
\label{nondual} \ee
then the \Lu\ measurement of $A$ does not disturb the measurement
of $B$. In other words, the expectation value of $B$ with respect
to {\em any} density operator $\rho$ is not affected by measuring
$A$. Introduce the {\em dual} \Lu\ map $\Lambda^D$, acting on
operators defined on ${\cal H}$, by
\be
\tr{\left[\Lambda(\rho) B\right]} = \tr{\left[\rho
\Lambda^D(B)\right]} \, .
\label{dual} \ee
Since Eq. (\ref{nondual}) is supposed to hold for any $\rho$, one
must have
\be
\Lambda^D (B ) = B \, ,
\label{dual2} \ee
which, after dropping the superscript, is the discrete counterpart
of Eq. (\ref{inv}). Now we can state \Lu ' theorem:
\be
\Lambda \left( B\right) =B \quad  \Leftrightarrow  \quad \left[B,
E_{i}\right] = 0\, , \quad  \mbox{ for all } i=1,2,\dots \, ,
\label{LT} \ee
{\em i.e.}, it is necessary and sufficient for $A=\sum_i^N a_i
E_i$ to commute with a (bounded) operator $B$ if the measurement
of $A$ should not disturb any measurement of $B$.

Originally, this theorem has been shown to hold for orthonormal
projections \cite{lueders51}; after a generalization to some
discrete POVMs had been obtained \cite{busch+98}, the theorem was
expected to hold under very general conditions. However, the
existence of a non-intuitive counterexample has been proved
non-constructively in \cite{gudder+02}. It is our purpose to
extend the validity of \Lu ' theorem to {\em continuous} POVMs
which are associated with coherent states on Lie groups.
\subsection*{\sf Outline and Summary}
In the following, we will consider POVMs which consist of
continuous families of one-dimensional projections onto coherent
states, or CS-POVMs, for short. The CS-POVMs for a spin and for a
particle provide well-known examples, being associated with the
group $SU(2)$ and the Heisenberg-Weyl group $H_3$, respectively.
However, coherent states can be defined for general Lie groups $G$
while retaining many of their properties. We will begin to discuss
the {\em L\"uders} map in general terms and specialize to
particular groups only later.

When considering {\em L\"uders}' map generated by coherent states
of an arbitrary (simple and simple connected) Lie group $G$, a
first general observation is that
\begin{itemize}
    \item the $P$- and the $Q$-symbol of a {\em L\"uders} operator
    coincide for the CS-POVM associated with a Lie group $G$.
\end{itemize}
Second, a simple form of this constraint is derived by expanding
the symbol of the operator in terms of harmonic functions
associated with the group $G$. The resulting condition on the
expansion coefficients will be shown to imply that
\begin{itemize}
  \item for the CS-POVM of a {\em spin} only the identity
        operator is {\em L\"uders};
  \item for the CS-POVM of a {\em particle} a countable
  family of linearly independent, unbounded {\em L\"uders} operators
  exists none of which commutes with the elements of the POVM.
\end{itemize}
Thus, for both the groups $SU(2)$ and $H_3$, the identity is found
to be the only {\em bounded} {\em L\"uders} operator which
commutes with the elements of the corresponding CS-POVM:
consequently, \Lu ' theorem also applies to these CS-POVMs.

Finally, it will be shown that the \Lu\ map implements anti-normal
ordering for operators which can be written as power series of
particle annihilation and creation operators.
\section*{\sf 2. {\em L\"uders} theorem for POVMs of coherent-states}
\subsection*{\sf Coherent States on Lie Groups and Harmonic Functions}
Given any finite-dimensional (simple and simply connected) Lie
group $G$, there is a canonical way to introduce coherent states
$\ket{\Omega}$ labelled by the points $\Omega$ of a well-defined
manifold $\mathbb{X}$. To do so, consider a unitary irreducible
representation $T(g)$ on a Hilbert space $\cal H$ of the elements
$g \in G$. Following closely the presentation given in
\cite{brif+99}, we choose a reference (or fiducial) state
$\ket{\psi_0}$ and define the set of coherent states by
\be
\ket{\psi_g} = T(g) \ket{\psi_0}\, , \quad g \in G \, .
\label{CSgen} \ee
Up to a phase, the reference state is left invariant by the
elements $h$ of the isotropy subgroup $H \subset G$,
\be
T(h) \ket{\psi_0} = e^{i\phi(h)} \ket{\psi_0}\, ,
               \quad h \in H \subset G \, .
\label{iso} \ee
Therefore, each group element can be written as as product
\be
g = \Omega h \, , \quad \Omega \in \mathbb{X} = G/H \, ,
                  \quad h \in H \, ,
\label{decomposeG} \ee
where $\mathbb{X}$ is the coset space obtained from dividing $G$
by its subgroup $H$. As the phase of a state has no physical
relevance, the set of coherent states is in a one-to-one
correspondence with the points $\Omega(g)$ of the manifold
$\mathbb{X}$. This suggests to denote coherent states by
$\ket{\Omega} \equiv \ket{\psi_\Omega}$.

A fundamental property of the coherent states $\ket{\Omega}$ is
their completeness in Hilbert space $\cal H$,
\begin{equation} \label{complomega}
\int_\mathbb{X} \,  d \mu (\Omega) \,
                   \ket{\Omega}\bra{\Omega} = I \, ,
\end{equation}
where integration is over the coset space $\mathbb{X}$ with
invariant measure $d \mu (\Omega)$, and $I$ is the identity in
$\cal H$.

Coherent states $\ket{\Omega}$ can be used to define symbolic
representations of operators, {\em i.e.} $c$-number valued
functions on the manifold $\mathbb{X}$ which can be understood as
the phase space of a classical system associated with the Lie
group $G$ \cite{perelomov86}. The $Q$-symbol of an operator $B$
acting in Hilbert space $\cal H$ is given by its expectation value
in coherent states,
\be
Q_B (\Omega)
    = \bra{\Omega} B \ket{\Omega} \, , \quad \Omega \in \mathbb{X}\, ;
\label{generalQ} \ee
due to analyticity properties of $Q_B (\Omega)$, these `diagonal'
matrix elements are sufficient to uniquely determine the operator
$B$. The $P$-symbol of $B$ \cite{glauber63,sudarshan63} arises if
one expresses $B$ as a linear combination of projection operators
$\ket{\Omega}\bra{\Omega}$:
\be
B = \int_\mathbb{X} d \mu (\Omega) \,
           P_B (\Omega) \, \ket{\Omega}\bra{\Omega} \, .
\label{generalP} \ee
The existence of the symbols $Q_B(\Omega)$ and $P_B(\Omega)$
depends in a subtle way on the properties of the operator $B$
\cite{perelomov86} but they are unique whenever they exist.
Furthermore, one can think of the symbols $Q_A (\Omega)$ and $P_A
(\Omega)$ as being dual to each other (cf. \cite{perelomov86}),
and, at least for particle coherent-states, they are related to
normal and anti-normal ordering of creation and annihilation
operators \cite{perelomov86,wilcox67}.

It is useful to introduce the harmonic functions $Y_\nu(\Omega)$
associated with the manifold $\mathbb{X}$ and, hence, with the
group $G$. Consider the Hilbert space $L^2(\mathbb{X},\mu)$ of
square integrable functions $u(\Omega)$ on the manifold
$\mathbb{X}$, with integration measure $d\mu(\Omega)$. The
eigenfunctions $Y_\nu(\Omega)$ of the Laplace-Beltrami operator on
$\mathbb{X}$ \cite{barut+80} constitute a complete orthonormal set
of functions in $L^2(\mathbb{X},\mu)$ since they satisfy
\be
\sum_\nu Y^*_\nu (\Omega) Y_\nu(\Omega^\prime)
   = \delta(\Omega - \Omega^\prime) \, ,
\label{compHarm} \ee
the right-hand-side being a delta function with respect to the
measure $\mu(\Omega)$, as well as
\be
\int_\mathbb{X} d \mu(\Omega)
         \, Y^*_\nu(\Omega) Y_{\nu^\prime}(\Omega)
           = \delta_{\nu \nu^\prime} \, .
\label{orthoHarm} \ee
Depending on the manifold $\mathbb{X}$ being compact or not, the
right-hand-side of (\ref{orthoHarm}) must be understood as a
Kronecker-delta or a Dirac-delta function (or suitable
combinations thereof). There is a simple expression for the
(modulus of) the overlap of two coherent states in terms of
harmonic functions:
\begin{equation} \label{skpromega}
| \bra{\Omega'} \Omega \rangle |^2
  = \sum_\nu \tau_\nu Y_\nu (\Omega') Y_\nu^* (\Omega) \, , \quad
  \tau_\nu \in \mathbf{R} \, ,
\end{equation}
where the numbers or functions $\tau_\nu$ depend on the actual
group.

\subsection*{\sf {\em L\"uders} map for CS-POVMs}
It is straightforward to generalize the {\em L\"uders} map
(\ref{general}) to POVMs which are continuous with respect to a
positive measure $\mu$. Let $\left(\Omega_0,\Sigma ,\mu \right) $
be a measure space, where $\Omega_0$ is a topological space with a
$\sigma-$algebra $\Sigma$ of subset of $\Omega_0$. Assume that,
for the Hilbert space ${\cal H} = L_2(\Omega_0,\mu)$, there is a
continuous map of the points points $\omega \in \Omega_0$ to the
set of positive linear operators $L\left( \mathcal{H}\right)$:
$\omega \mapsto E_{\omega }\geq 0$. If the operators $E_\omega$
provide, in addition, a resolution of unity,
\be
\int_{\Omega_0} d\mu (\omega) \, E_{\omega} = I \, ,
\label{resolution1} \ee
then the operators
\begin{equation} \label{generalPOVM}
E ( \sigma )
 = \int_{\sigma} d\mu (\omega) \, E_{\omega} \, , \quad
  \sigma \in \Sigma \, .
\end{equation}
define a continuous POVM. It is natural to associate with it a
{\em L\"uders} map $\Lambda (B)$ of an operator $B$ by defining
\begin{equation}
\Lambda \left( B \right)
  = \int_{\Omega} \, d\mu \left( \omega \right) \,  E_{\omega }^{1/2}
   \, B \, E_{\omega }^{1/2} \, ,
\label{generalLueders} \end{equation}
which is a unital, completely positive linear map on
$L\left(\mathcal{H}\right)$. Due to the completeness relation
(\ref{complomega}), the self-adjoint coherent-state projectors
\begin{equation} \label{povmelementsomega}
E_\Omega \equiv \ket{\Omega} \bra{\Omega}
         = E^{1/2}_\Omega \, ,
                            \quad \Omega \in \mathbb{X} \, ,
\end{equation}
are seen to define a POVM in the sense just described.

Any operator $B$ defined on $L_2(\mathbb{X})$ is {\em L\"uders}
with respect to the CS-POVM $E_\Omega, \Omega \in \mathbb{X}$, if
it satisfies the relation $B = \Lambda (B)$ with $E_\omega$ in
(\ref{generalLueders}) replaced by $E_\Omega$,
\begin{equation}\label{OmegaLueders}
B = \int_\mathbb{X} \, d \mu (\Omega) \,
    \ket{\Omega} \bra{\Omega} B \ket{\Omega} \bra{\Omega}
  = \int_\mathbb{X} \, d \mu (\Omega) \,
    Q_B(\Omega) \ket{\Omega} \bra{\Omega}\, .
\end{equation}
Upon comparing this equation with (\ref{generalP}), we observe
that the {\em L\"uders} property has, for any CS-POVM, the
following general interpretation: an operator $B$ is {\em
L\"uders} if and only if its $P$- and $Q$-symbols coincide,
\be
P_B(\Omega) = Q_B(\Omega) \, .
\label{eqsymb} \ee
To the best of our knowledge, this set of operators---which we
will call {\em well-ordered}---has not been introduced before.

The constraint (\ref{OmegaLueders}) takes a particularly simple
form upon expanding the $Q$-symbol of $B$ in harmonic functions,
\be
Q_B (\Omega) = \sum_\nu B_\nu Y_\nu (\Omega)  \, ,
\label{expandQ_B} \ee
which is possible according to (\ref{compHarm}). The expansion
coefficients are given by
\begin{equation} \label{Bxicoeffsomega}
B_\nu = \int_\mathbb{X} \,  d\mu (\Omega) \,
           Q_B (\Omega) Y_\nu^* (\Omega) \, .
\end{equation}
Take the expectation value of (\ref{OmegaLueders}) in the coherent
state $\ket{\Omega'}$ and use the relation (\ref{skpromega}) for
the overlap $| \bra{\Omega'} \Omega \rangle |^2$. This leads to
\be \label{newexprBomega}
Q_B (\Omega') = \sum_\nu \tau_\nu
   \left[ \int_\mathbb{X} \, d\mu (\Omega) Q_B(\Omega) Y_\nu^*
   (\Omega)\right] Y_\nu (\Omega')
 = \sum_\nu \tau_\nu B_\nu Y_\nu (\Omega') \, ,
\ee
where (\ref{Bxicoeffsomega}) has been used. Uniqueness of the
expansion (\ref{expandQ_B}) implies that the coefficients of a {\em L\"uders} operator must satisfy the condition
\begin{equation} \label{POVMconditionomega}
B_\nu = \tau_\nu B_\nu \, , \quad \mbox{ for all } \nu \, .
\end{equation}
As mentioned above, the actual form of the quantities $\tau_\nu$
depend on the group $G$ under consideration. To proceed, we
therefore need to specify the system of coherent states we work
with, that is, the group $G$. Explicit conclusions about {\em
L\"uders} operators for CS-POVMs will be derived now for the
groups $SU(2)$ and $H_3$.

\section*{\sf 3. {\em L\"uders} operators for the CS-POVM of a spin}
Consider a Hilbert space ${\cal H}_s$ of dimension $(2s+1)$,
carrying an irreducible representation of the group $G=SU(2)$.
Each space ${\cal H}_s$ is associated with a spin of length $s \in
\{ 1/2, 1, 3/2, \ldots \}$. To introduce spin-coherent states, it
is convenient to select states of highest (lowest) weight
$\ket{\pm s}$ as reference states (cf.
\cite{perelomov86,arecchi+72}. These states are invariant under a
change of phase, hence the isotropy group is given by $H=U(1)$.
Therefore, the coset space is the surface of a sphere: $\mathbb{X}
= SU(2)/U(1) = {\cal S}^2$, which corresponds to the phase space
of a classical spin.

The resolution of unity $I$ in ${\cal H}_s$ using  spin-coherent
states $\ket{\bf n}$ reads
\begin{equation} \label{unity}
I = \int_{{\cal S}^2}  d \mu({\bf n}) \,
              \ket{\bf n} \bra{\bf n} \, , \qquad
              d\mu({\bf n}) = \frac{2s+1}{4\pi}
                        \sin \vartheta d\vartheta \, d \phi \, ,
\end{equation}
where each unit vector ${\bf n} \in {\sf I \! R}^3$ denotes a
point with spherical coordinates $(\vartheta,\varphi)$, located on
the unit sphere ${\cal S}^2$. The continuous family of operators
\begin{equation} \label{Eops}
E_{\bf n} = \ket{\bf n} \bra{\bf n} \, , \quad \mbox{ with } \quad
I = \int_{{\cal S}^2}  d\mu({\bf n}) \, E_{\bf n} \, ,
\end{equation}
defines the CS-POVM of $SU(2)$. Being a projector, the positive
square root of each operator $E_{\bf n}$ is equal to itself:
$E^{1/2}_{\bf n} = \ket{\bf n} \bra{\bf n}$. Therefore, a
self-adjoint operator $B \in L({\cal H}_s)$ is {\em L\"uders} with
respect to the POVM (\ref{Eops}) if
\begin{equation} \label{Blueders}
B = \int_{{\cal S}^2}  d\mu({\bf n})
          \ket{\bf n} \bra{\bf n} B \ket{\bf n}  \bra{\bf n}
   \equiv \int_{{\cal S}^2}  d\mu({\bf n}) Q_B ({\bf n})
                                   \ket{\bf n} \bra{\bf n} \, .
\end{equation}
Following the strategy outlined earlier, we will show now that any
operator $B$ satisfying (\ref{Blueders}) must be a real multiple
of unity: $B = \lambda I$, so that $B$ commutes with all elements
of the CS-POVM for a spin,
\begin{equation} \label{finalresult}
 \left[ B , E_{\bf n}  \right] = 0 \, , \quad {\bf n} \in {\cal S}^2 \,
.
\end{equation}

Consider the expectation value of Eq. (\ref{Blueders}) in the
coherent state $\ket{{\bf n}^\prime}$,
\begin{equation} \label{expB}
Q_B ({\bf n}^\prime)
   = \int_{{\cal S}^2}  d\mu({\bf n}) \,
    Q_B({\bf n}) |\bra{{\bf n}} {\bf n}^\prime \rangle |^2 \, .
\end{equation}
The function $Q_B ({\bf n})$, the Q-symbol of the operator $B$, is
smooth on the sphere ${\cal S}^2$, and it can be written as a
linear combination of $(2s+1)^2$ spherical harmonics $Y_{lm}({\bf
n})$,
\begin{equation} \label{expandq}
Q_B({\bf n})  = \sqrt{\frac{4\pi}{2s+1}}
      \sum_{l=0}^{2s} \sum_{m=-l}^l B_{lm} Y_{lm} ({\bf
n}) \, ,
\end{equation}
with expansion coefficients
\begin{equation} \label{expcoeff}
B_{lm} = \sqrt{\frac{4\pi}{2s+1}}
           \int_{{\cal S}^2}  d\mu({\bf n}) \,
                  Q_B ({\bf n}) \, Y_{lm}^* ({\bf n})\, .
\end{equation}
Note that these expressions are connected to the general formulas
through identifying $Y_\nu (\Omega) \leftrightarrow
\sqrt{4\pi/(2s+1)} Y_{lm}(\bf n)$. Rewrite the scalar product
(\ref{expB}) by means of the
addition theorem for spherical harmonics, %\cite{varilly+89},
\bea \label{scpr}
| \bra{{\bf n}} {\bf n}^\prime \rangle |^2 &=& \left( \frac{1+{\bf n}\cdot {\bf n}^\prime}{2} \right)^{2s} \nonumber \\
&=& \sum_{l=0}^{2s} \frac{2l+1}{2s+1}
        {\clebsch{s}{l}{s}{s}{0}{s}}^2 P_l({\bf n}\cdot {\bf n}^\prime) \nonumber \\
&=& \frac{4\pi}{2s+1} \sum_{l=0}^{2s} \sum_{m=-l}^{l}
         {\clebsch{s}{l}{s}{s}{0}{s}}^2 Y_{lm}^* ({\bf n}) Y_{lm} ({\bf n}^\prime) \, ,
\eea
where the functions $P_l(x)$ are the Legendre polynomials. Upon
inserting (\ref{expandq}) and (\ref{scpr}), integration of the
right-hand-side of Eq. (\ref{expB}) gives (after replacing ${\bf
n}^\prime$ by ${\bf n}$)
\begin{equation} \label{diffB}
Q_B ({\bf n})
 =\sqrt{\frac{4\pi}{2s+1}} \sum_{l=0}^{2s} \sum_{m=-l}^l
     {\clebsch{s}{l}{s}{s}{0}{s}}^2 B_{lm} Y_{lm}
                                     ({\bf n}) \, .
\end{equation}
This expansion and Eq. (\ref{expandq}) can only hold
simultaneously if the coefficients of the harmonics satisfy
\begin{equation} \label{condcoeff}
B_{lm} = {\clebsch{s}{l}{s}{s}{0}{s}}^2 B_{lm} \, ,
\end{equation}
which is (\ref{POVMconditionomega}) for the group $SU(2)$.  The
$m$-independent Clebsch-Gordan coefficients correspond to the
numbers $\tau_\nu$ introduced in (\ref{skpromega}), and  they take
values
\begin{equation} \label{cg}
 {\clebsch{s}{l}{s}{s}{0}{s}}^2 = \frac{(2s)! (2s+1)!}{(2s-l)! (2s+1+ l)!} \, .
\end{equation}
Since
\begin{equation} \label{clebschvalues}
 {\clebsch{s}{0}{s}{s}{0}{s}} = 1\, , \qquad
  0 < {\clebsch{s}{l}{s}{s}{0}{s}} < 1 \, , \quad
  l = 1, 2, \ldots , 2s,
\end{equation}
the coefficients $B_{lm}$ with $l\neq 0$ in (\ref{condcoeff}) must
vanish; thus, the expansion (\ref{expandq}) of a {\em L\"uders}
operator satisfying (\ref{Blueders}) contains only one nonzero
term, $B_{00}$, and $B$ is proportional to $Y_{00}({\bf n})$, {\em
i.e.}, the identity. Hence, it commutes with any operator,
including the set $E_{\bf n}$, so that Eq. (\ref{finalresult})
follows. At the same time we have shown that the identity is the
only operator in ${\cal H}_s$ such that its $Q$- and $P$-symbol
coincide.
\section*{\sf 4. {\em L\"uders} operators for the CS-POVM of a particle}
The kinematics of a quantum particle on the real line $\mathbb{R}$
is described by the creation and annihilation operators $a$ and
its adjoint $a^\dagger$ which satisfy $[ a , a^ \dagger ] = I$.
The operators $a$, $a^\dagger$, and the identity $I$ generate the
Heisenberg-Weyl algebra $h_3$; finite transformations, that is,
elements of the {\em group} $H_3$, are given by the phase-space
displacement or shift operators
\begin{equation} \label{displacement}
D(\alpha) = \exp [ \alpha a^\dagger - \alpha^* a ] \, , \quad
            \alpha \in \mathbb{C} \, .
\end{equation}
In fact, they provide an irreducible projective representation of
the group $H_3$ in $L_2(\mathbb{R})$,
\begin{equation} \label{H3irrep}
D(\alpha) D(\alpha')
          = \exp \left[ \frac{i}{2} \left( \alpha \alpha^{\prime*}
                        - \alpha^* \alpha' \right) I \right]
          D(\alpha + \alpha') \, .
\end{equation}
The (overcomplete) family of coherent states $\ket{\alpha}$ in the
Hilbert space $L_2(\mathbb{R})$ is obtained by displacing the
fiducial state $\ket{0}$, say, with $a \ket{0}=0$, by arbitrary
amounts $\alpha \in \mathbb{C}$:
\begin{equation}
\ket{\alpha} = D(\alpha) \ket{0} \, .
\end{equation}
The isotropy subgroup of $H_3$ is again isomorphic to $U(1) \sim
\exp [ i \gamma I ], \gamma \in [0,2\pi)$, so that the manifold
labeling coherent states is given by the complex plane $\mathbb{X}
= H_3/ U(1) = \mathbb{C}$, corresponding indeed to the phase space
of a classical particle on the real line.

The completeness relation for the particle-coherent states
$\ket{\alpha}$ reads
\begin{equation} \label{complpart}
I = \int_\mathbb{C} d \mu (\alpha) \,
           \ket{\alpha} \bra{\alpha}\, , \quad
             d \mu (\alpha) = \frac{1}{\pi} d^2 \alpha \, ,
\end{equation}
and it can be understood as defining a POVM for the continuous
family of projection operators
\begin{equation} \label{povmelementspart}
E_\alpha = \ket{\alpha} \bra{\alpha} = E^{1/2}_\alpha \, ,
                            \quad \alpha \in \mathbb{C} \, .
\end{equation}
The operator $B$ on $L_2(\mathbb{R})$ is {\em L\"uders} with
respect to the POVM $ E_\alpha, \alpha \in \mathbb{C}$, if it is
invariant under the {\em L\"uders} map $B \mapsto \Lambda (B)$,
{\em i.e.},
\begin{equation}\label{particleLueders}
B = \int_\mathbb{C} \, d \mu (\alpha) \,
     \ket{\alpha}\bra{\alpha} B \ket{\alpha} \bra{\alpha}
   = \int_\mathbb{C} \, d \mu (\alpha) \,
     Q_B (\alpha) \ket{\alpha} \bra{\alpha}  \, ,
\end{equation}
where $\bra{\alpha} B \ket{\alpha} = Q_B(\alpha)$ is the
$Q$-symbol of the operator $B$. As shown, this relation forces the
$Q$-symbol of a {\em L\"uders} operator to coincide with its
$P$-symbol,
\begin{equation}\label{particleLuedersP}
B =  \frac{1}{\pi} \int_\mathbb{C} \, d \mu (\alpha)
    P(\alpha) \ket{\alpha}\bra{\alpha}  \, ,
\end{equation}
if it exists.

We will now search for {\em bounded} {\em L\"uders} operators $B$
which commute the members $E_\alpha$ of the CS-POVM
(\ref{complpart}) for a particle. We begin to look at simple
examples of {\em L\"uders} operators, followed by a systematic
construction of all well-ordered {\em L\"uders} operators. In
addition to the identity, a countable family of {\em unbounded},
linearly independent {\em L\"uders} operators will emerge none of
which commutes with the elements of the CS-POVM. Finally, an
unexpected relation of the {\em L\"uders} map to operator
orderings is established for particle coherent states.

\subsection*{\sf Examples of unbounded {\em L\"uders} operators}
It is straightforward to apply the map $\Lambda$ to unbounded operators such as position $Q=(a+a^\dagger)/2$ and momentum $P=(a-a^\dagger)/2i$. Using the equation $a \ket{\alpha} = \alpha \ket{\alpha}$ and its adjoint implies that
\begin{eqnarray}\label{LuedersinvQP}
\Lambda \left( Q \right)
 & = & \int_{\mathbb{C}} d\mu (\alpha) \,  \ket{\alpha}
        \bra{\alpha} Q \ket{\alpha}\bra{\alpha} \,
    =    \int_{\mathbb{C}} d\mu (\alpha) \,
       \frac{1}{2}(\alpha + \alpha^*) \ket{\alpha}
         \bra{\alpha}   \nonumber \\
 & = & \frac{1}{2} \int_{\mathbb{C}} d\mu (\alpha) \,
       a \ket{\alpha} \bra{\alpha}
       + \frac{1}{2} \int_{\mathbb{C}} d\mu (\alpha) \,
        \ket{\alpha} \bra{\alpha} a^\dagger
       = Q \, ,
\end{eqnarray}
and similarly
\be
\Lambda ( P )  = P \, .
\label{mapP} \ee
While being invariant under $\Lambda$, the operators $Q$ and $P$
are neither positive nor bounded, and they do not commute with the
projectors $E_\alpha$ since the expectation value of the
commutator in the coherent state $\ket{\beta}$ is, in general,
different from zero:
\be
\bra{\beta} \left[ Q , E_\alpha \right] \ket{\beta}
  = \frac{1}{2} \left( (\alpha- \alpha^*) - (\beta -\beta^*)
  \right)  |\bra{\alpha} \beta \rangle|^2 \, .
\label{nonzereocomm} \ee
Using the relation $D^\dagger(\alpha) a D(\alpha) = a-\alpha$, its
adjoint, and the commutation relations of $a$ and $a^\dagger$, one
shows that {\em L\"uders}' map acts on the operators $Q^2$ and
$P^2$ according to
\bea\label{LuedersinvQQ}
\Lambda ( Q^2 )
    &=& Q^2 + 2\bra{0} Q^2 \ket{0} I =
      Q^2 + \frac{1}{2} I \, , \nonumber \\
\Lambda ( P^2 )
    &=& P^2 + 2\bra{0} P^2 \ket{0} I =
      P^2 + \frac{1}{2} I \, .
\end{eqnarray}
Consequently, appropriate quadratic combinations of position and
momentum turn out to be {\em L\"uders},
\begin{equation}\label{lincombpow}
\Lambda_{\Gamma} \left( Q^2 -P^2 \right) = Q^2 -P^2 \, .
\end{equation}
However, this indefinite, unbounded operator does not commute with
all projections $E_\alpha$ as follows from $\bra{0}[ Q^2 -P^2,
E_\alpha ] \, \ket{0}$ $=$ $(\alpha^2- \alpha^{*2}) \, |\bra{0}
\alpha\rangle|^2$, for example. In the next section a family of
similar {\em L\"uders} operators will be constructed.
\subsection*{\sf Construction of {\em L\"uders} operators}
Let us turn now to the problem of finding all operators which are
{\em L\"uders} with respect to the CS-POVM $E_\alpha$ of a
particle. {i.e.} all well-ordered operators. The argument will
resemble the one given in the case of a spin.

Expand the $Q$-symbol of an operator $B$ as
\begin{equation} \label{expandB}
Q_B (\alpha) = \int_\mathbb{C} \, d\mu (\xi) \,
  B_\xi \exp \left[ \alpha \xi^* - \alpha^* \xi \right] \,
\end{equation}
where the coefficients $B_\xi$ are given by
\begin{equation} \label{Bxicoeffs}
B_\xi = \int_\mathbb{C} \,  d\mu (\alpha) \, Q_B (\alpha)
        \exp \left[ -\left( \alpha \xi^* - \alpha^* \xi \right)
         \right] \, .
\end{equation}
Here, the functions $\exp \left[ \alpha \xi^* - \alpha^* \xi
\right]$ are the complete orthonormal set of harmonic functions in
the complex plane, corresponding to $Y_\nu(\Omega)$. Since the
$Q$-symbol of a hermitean operator is real, $Q_B (\alpha)=
\bra{\alpha} B \ket{\alpha}^* = Q_B^* (\alpha)$, the coefficients
must satisfy the relation
\begin{eqnarray} \label{Bxicoeffscond}
B_\xi^* &=& \int_\mathbb{C} \,  d\mu (\alpha) \, Q_B^*(\alpha)
        \exp \left[ -\left( \alpha^* \xi - \alpha \xi^* \right)
         \right] \nonumber \\
        &=& \int_\mathbb{C} \,  d\mu (\alpha) \, Q_B(\alpha)
        \exp \left[ -\left( \alpha (-\xi)^* - \alpha^* (-\xi) \right)
         \right]
         = B_{-\xi} \, .
\end{eqnarray}
We will turn (\ref{particleLueders}) into a condition for the
expansion coefficients $B_\xi$ of a {\em L\"uders} operator which
can be solved explicitly. Take the expectation value of the
operator $B$ in (\ref{particleLueders}) in the coherent state
$\ket{\beta}$, and use the identity
\begin{eqnarray} \label{skpr}
| \bra{\alpha} \beta \rangle |^2
  &=& \exp \left[ - | \alpha - \beta |^2 \right]
  \nonumber \\
  &=& \int_\mathbb{C} \, d\mu (\xi) \,
    e^{-\xi \xi^*} \exp \left[ \beta \xi^* - \beta^* \xi  \right] \,
                  \exp \left[ - \alpha \xi^* + \alpha^* \xi \right]
                  \, ,
\end{eqnarray}
leading to
\begin{eqnarray} \label{newexprB}
Q_B (\beta) &=& \int_\mathbb{C} \, d\mu (\xi) \,
  e^{-\xi \xi^*} \left[ \int_\mathbb{C} \, d\mu (\alpha) \,  Q_B (\alpha)
  \exp \left[ - (\alpha \xi^* - \alpha^* \xi) \right] \right]
  \exp \left[ \beta \xi^* - \beta^* \xi \right] \, , \nonumber \\
&=& \int_\mathbb{C} \, d\mu (\xi) \,
            e^{-\xi \xi^*} B_\xi
            \exp \left[ \beta \xi^* - \beta^* \xi \right] \, ,
\end{eqnarray}
where (\ref{Bxicoeffs}) has been used. Due to the uniqueness of
the expansion (\ref{expandB}), the expansion coefficients of any
{\em L\"uders} operators must satisfy
\begin{equation} \label{POVMcondition}
B_\xi = e^{- \xi \xi^*} B_\xi \, ,
\end{equation}
which is the equivalent of (\ref{condcoeff}) for continuous
variables. Consequently, the coefficients $B_\xi$ are necessarily
zero for all values of $\xi$ except $\xi=0$, and there are no
solutions in terms of ordinary functions. If allowing for
generalized functions, $B_\xi$ is necessarily a distribution of
finite order \cite{bremermann65}, that is, a linear combination of
a $\delta$-distribution and finite derivatives of it,
\begin{equation} \label{zerosupport}
B_\xi = \sum_{n+m=0}^N b_{nm} \partial_\xi^n
      \partial_{\xi^*}^m \delta(\xi) \, , \quad b_{nm} \in \mathbb{C}
      \, , \quad n,m = 0,1,2, \ldots \, , \quad
      N = 0,1,2, \ldots
\end{equation}
The function $B_\xi$ must satisfy (\ref{Bxicoeffscond}) leading to
\begin{equation} \label{reality}
b_{nm} = (-)^{m+n} b^*_{mn} \, , \quad n,m = 0,1,2,\ldots \, ,
\end{equation}
and the $\delta (\xi)$-function is real,
\begin{equation} \label{complexdelta}
\delta (\xi) =
%\frac{1}{\pi}
\int_\mathbb{C} \, d\mu (\alpha)
  \exp \left[ \alpha \xi^* - \alpha^* \xi \right]
          = \delta(-\xi) = \delta^* (\xi) \, .
\end{equation}
Only some of the distributions (\ref{zerosupport}) will satisfy
(\ref{POVMcondition}) since one must have
\begin{equation}\label{condA}
Q_B (\alpha) = \int_\mathbb{C} \, d\mu (\xi)
       \left[ D_{N} \delta(\xi) \right]
       e^{- \xi \xi^*} e^{\alpha \xi^* - \alpha^* \xi}
    = \int_\mathbb{C} \, d\mu (\xi)
    \left[ D_{N} \delta(\xi) \right]
         e^{\alpha \xi^* - \alpha^* \xi} \, ,
\end{equation}
where
\be
D_{N} = \sum_{n+m=0}^N b_{nm} \partial_\xi^n
      \partial_{\xi^*}^m \, .
\label{diffop} \ee
Partial integrations in (\ref{condA}) lead to the requirement
\begin{equation}\label{condB}
 \left[ D^{\dagger}_{N}
       e^{- \xi \xi^*} e^{\alpha \xi^* - \alpha^* \xi}
       \right]_{\xi=\xi^*= 0}
    = \left[ D^{\dagger}_{N}
         e^{\alpha \xi^* - \alpha^* \xi} \right]_{\xi=\xi^* = 0} \, ,
\end{equation}
where the adjoint $D^{\dagger}_N$ of $D_{N}$ is obtained from
replacing $b_{nm}$ by $(-)^{n+m} b_{nm}$ in (\ref{diffop}). It is
shown in the Appendix that this condition is satisfied if and only
if
\be
b_{nm} = 0 \, ,  \quad 1 \leq m,n  \leq N \, ,
\label{zerocoeff} \ee
{\em i.e.}, only terms with at least one index (that is, $m$ or
$n$ or both) equal to zero will contribute to the symbol of a
well-ordered operator. Therefore, only coefficients of the form
\begin{equation}\label{Qlueders}
B_\xi = \sum_{n=0}^N \left( b_{n0} \partial_{\xi}^n
                     + (-)^nb_{n0}^* \partial_{\xi^*}^n \right)
                     \delta(\xi)
\end{equation}
which, upon partial integration in (\ref{expandB}), give rise to
$Q$-symbols of \Lu\ operators, 
\be
Q_B (\alpha) = \sum_{n=0}^N \left( b_{n0} \alpha^{*n}
                     + b_{n0}^* \alpha^n \right) \, .
\label{invsymb} \ee
The operators corresponding to these symbols are given by
\be
B = b_{0}I + \sum_{n=1}^N
 \left( b_n^q B^q_n + b_n^p B^p_n \right) \, ,
\label{allparticleluedersops} \ee
a linear combination of the identity and $2N$ hermitean operators
\be
B_n^q = \frac{1}{2} \left(  a^n + a^{\dagger \, n} \right) \quad
 \mbox{ and } \quad
 B_n^p = \frac{1}{2i} \left( a^n - a^{\dagger \,n} \right) \, , \quad
 n = 1,2, \ldots , N \, ,
\label{luedersbasis} \ee
which satisfy (\ref{particleLueders}), and $(2N+1)$ real
coefficients
\be
b_0 = 2b_{00} \, , \quad
   b^q_n = b_{n0} + b_{n0}^* \, , \quad
   b^p_n = \frac{1}{i} \left( b_{n0} - b_{n0}^* \right) \, ,
   \quad n= 1,2,\ldots , N \, .
\label{invops} \ee
%
%Consequently, the set of all \Lu\ operators
%(\ref{allparticleluedersops}) can be parameterized by $(2N+1)$
%real parameters.
For $N=2$, for example, it follows that not only the operators
$Q,P$, and $Q^2 - P^2$ are {\em L\"uders} but also
\be
B_2^p = \frac{1}{2i} \left(  a^2 - a^{\dagger \, 2} \right)
\propto QP+PQ \, .
\label{newlue} \ee
Every bounded \Lu\ operator is necessarily a multiple of the
identity.
\subsection*{\sf {\em L\"uders} map and operator ordering}
It is easy to understand why the operators $B_n, n= 1,2,
\ldots,N$, in (\ref{invops}) are {\em L\"uders}. Consider any
hermitean operator $B $ given as a finite polynomial in $a$ and
$a^\dagger$. Using their commutation relation, one can bring the
annihilation operators either to the right or to the left,
\begin{equation}\label{orders}
B (a , a^\dagger)
  = \sum_{m,n} \beta^{\cal N}_{nm}
           a^{\dagger \, m} a^n
  = \sum_{m,n} \beta^{\cal A}_{nm}
           a^m  a^{\dagger \, n} \, ,
\end{equation}
corresponding to normal- and anti-normal ordering of $B$,
respectively \cite{berezin80}. It is straightforward to calculate
the {\em L\"uders} transform of $B$ if it is written in normal
order:
\be\label{left}
\Lambda(B(a , a^\dagger))
 = \sum_{m,n} \beta^{\cal N}_{nm} \,
    \Lambda \left( a^{\dagger \, m} a^n \right)
 = \sum_{m,n} \beta^{\cal N}_{nm} a^n
                a^{\dagger \, m}  \, ,
\ee
since
\bea
\Lambda \left( a^{\dagger \, m} a^n \right)
  &=&  \int_\mathbb{C} \, d\mu (\alpha) \,
   \ket{\alpha}\bra{\alpha} a^{\dagger \, m} a^n
   \ket{\alpha}\bra{\alpha}
    = \int_\mathbb{C} \, d\mu (\alpha) \,
   \alpha^n \ket{\alpha}\bra{\alpha} \alpha^{*\, m}
\nonumber \\
&=&     a^n \left(\int_\mathbb{C} \, d\mu (\alpha) \,
\ket{\alpha}\bra{\alpha}\right) a^{\dagger \, m}
 = a^n a^{\dagger \, m} \, .
\label{lambdapower} \eea
Thus, the effect of $\Lambda$ is to push each creation operator
$a^\dagger$ to the right as if it would commute with the
annihilation operator $a$. In other words, the map $\Lambda$
provides an explicit form of the operator ${\cal A}$ which
generates anti-normal order of an operator \cite{wilcox67}. This
operator and its twin ${\cal N}$, which brings a given operator
into normal order, are useful tools to evaluate expectation values
or Baker-Campbell-Hausdorff relations, for example
\cite{wilcox67}.

To conclude: an operator $B$ is to be invariant under $\Lambda$,
the normally and anti-normally ordered forms of an operator $B$
must coincide,
\begin{equation}\label{ordercondition}
\sum_{m,n} \beta^{\cal N}_{nm} a^n
                 a^{\dagger \, m}
 = \sum_{m,n} \beta^{\cal A}_{nm} a^m
                a^{\dagger \, n} \, ,
\end{equation}
that is, $\beta^{\cal N}_{nm} = \beta^{\cal A}_{nm}$.  This is
obviously true for the linear combinations of powers of $a$ and
$a^\dagger$ given in (\ref{invops}), defining the family of
well-ordered operators.

\section*{\sf 5. Discussion}

We have shown that there is only one \Lu\ operator, the identity,
for the CS-POVM of $SU(2)$ while a family of $(2N+1)$ linearly
independent, unbounded, and well-ordered operators exists in the
case of $H_3$. It is plausible that our study exhausts all
possibilities which may arise for CS-POVMs of (simple and simply
connected) Lie groups: we expect only the identity as a \Lu\
operator for {\em compact} Lie groups such as $SU(N)$, and a
countable family for a CS-POVM associated with non-compact groups
such as $SU(N-n,n), 1 \leq n < N$. If we restrict our attention to
bounded operators, we expect \Lu ' theorem to hold for the CS-POVM
of any Lie group $G$.
\subsection*{\sf Appendix}
We will show here that any operator compatible with
(\ref{particleLueders}) must have a $Q$-symbol with expansion
coefficients given by
\begin{equation}\label{Qluedersapp}
B_\xi = \sum_{n=0}^N \left( b_{n0} \partial_{\xi}^n
                     + (-)^n b_{n0}^* \partial_{\xi^*}^n \right) \delta (\xi) \, ;
\end{equation}
this means, in particular, that most of the coefficients $b_{nm}$
are equal to zero:
\be
b_{nm} = 0 \, , \quad \mbox{ for } 1 \leq m,n \leq N \, .
\label{zerocoeff2} \ee
In a first step, evaluate the right-hand-side of (\ref{condB}):
\begin{equation}\label{condD2}
\left[ \sum_{n+m=0}^N (-)^{n+m} b_{nm} \partial_\xi^n
      \partial_{\xi^*}^m
      e^{\alpha \xi^* - \alpha^* \xi} \right]_{\xi=0}
   = \sum_{n+m=0}^N (-)^m b_{nm} \alpha^m \alpha^{* \, n} \, .
\end{equation}
To evaluate the left-hand side, use the relation
\begin{equation}\label{condC}
 \partial_\xi  \left(e^{- \xi \xi^*}  f(\xi ) \right)
   = e^{- \xi \xi^*} ( -\xi^* + \partial_\xi)f(\xi)
%   \quad \mbox{ and } \quad
%\partial_{\xi^*} \left( e^{- \xi \xi^*} g(\xi^*) \right)
%   = e^{- \xi \xi^*} ( - \xi + \partial_{\xi^*}) g(\xi^*) \, ,
%
\end{equation}
and its complex conjugate for any smooth function $f$. This leads
to
\begin{eqnarray}\label{condE}
\partial_\xi^n \partial_{\xi^*}^m e^{- \xi \xi^*}
&=& e^{- \xi \xi^*} ( -\xi^* + \partial_\xi    )^n
                  ( -\xi   + \partial_{\xi^*})^m  \\
&=& e^{- \xi \xi^*} \sum_{\nu=0}^n \sum_{\mu=0}^m
     \bc{n}{\nu} \bc{m}{\mu}
     \left( -\xi^* \right)^{n-\nu}
     \partial_\xi^\nu \left( -\xi \right)^\mu \partial_{\xi^*}^{m-\mu}
     \nonumber\, .
\end{eqnarray}
According to Eq. (\ref{condB}), these operators must be applied to
the function $e^{\alpha \xi^* - \alpha^* \xi}$. Each derivative
$\partial_{\xi^*}$ produces a factor $\alpha$, while the action of
the derivatives $\partial_{\xi}$ is more complicated:
\begin{eqnarray}\label{condF}
\partial_\xi^\nu
 \left(\left( -\xi \right)^\mu e^{\alpha \xi^* - \alpha^* \xi}
 \right)
 &=& \sum_{s=0}^\nu \bc{\nu}{s}
 \frac{\partial (-\xi)^\mu}{\partial \xi^s}
 \frac{\partial^{\nu-s} e^{\alpha \xi^* - \alpha^* \xi}}{\partial
                             \xi^{\nu-s}} \\
 &=& \sum_{s=0}^\nu \bc{\nu}{s} \frac{\mu!(-)^{s}}{(\mu-s)!}
     (-\xi)^{\mu-s} (-\alpha^*)^{\nu-s}
     e^{\alpha \xi^* - \alpha^* \xi} \nonumber  \, ;
\end{eqnarray}
due to $1/\Gamma(-k) = 0, k=0,1, 2, \ldots$, there are no
contributions to the sum if $s$ exceeds $\mu$. Now that the
derivatives have been evaluated, one can set $\xi=\xi^*=0$ in the
resulting expression: the terms with non-zero powers of $\xi$ or
$\xi^*$ vanish, so the sums simplify according to
\begin{equation}\label{deltas}
\left( -\xi \right)^{\mu-s} \to \delta_{\mu s}
    \quad \mbox{and} \quad
\left( -\xi^* \right)^{n-\nu} \to \delta_{n \nu} \, .
\end{equation}
The left-hand-side of (\ref{condB}) becomes
\begin{equation}\label{lhs}
\sum_{n+m=0}^N (-)^{m} b_{nm}
 \sum_{s=0}^{s_0} s! {\bc{m}{s}} {\bc{n}{s}}
  \alpha^{m-s}
 \alpha^{* \,  n-s} \,  ,
\end{equation}
where $ s_0 = \min (m,n) $. Note that the term with $s=0$ in this
expression is identical to the right-hand-side of (\ref{condD2})
which implies that the equality (\ref{condA}) is satisfied if
\begin{equation}\label{symbolscond}
\sum_{n+m=0}^N (-)^{m} b_{nm}
 \sum_{s=1}^{s_0} s! {\bc{m}{s}} {\bc{n}{s}}
 \alpha^{m-s} \alpha^{* \, n-s} = 0
\end{equation}
holds for all complex numbers $\alpha$. This equation does not
restrict the coefficients $b_{n0}, 0 \leq n \leq N$ and $b_{0m}, 0
\leq m \leq N$: if either $m$ or $n$ are equal to zero, the sum
over $s$ is empty since $s_0 = 0$. However, {\em all} other
coefficients must vanish as can be seen in the following way.
Writing $\alpha = r \exp [i\varphi]$, Eq. (\ref{symbolscond})
turns into a sum of terms multiplying phase factors $\exp [
i(m-n)\varphi] \equiv \exp [i k \varphi]$, $ k= 0,1, 2, \ldots ,
N-1$. Each of these terms must vanish individually due to the
linear independence of the exponentials. Their coefficients, in
turn, are power series in $r$ which can be shown to vanish
identically only if $b_{1N} = 0$ for $\exp [i(N-1) \varphi]$,
$b_{2N} = 0 \Rightarrow b_{1 \, N-2} =0$ for $\exp [i (N-2)
\varphi ]$, etc.  Taking into account that $b_{nm} = (-)^{m+n}
b^*_{nm}$, the coefficients $B_\xi$ of {\em L\"uders} operators
finally read
\begin{equation}\label{there we are!}
B_\xi = \left( \sum_{n=0}^N b_{n0} \partial_\xi^n
               + \sum_{m=0}^N b_{0m} \partial_{\xi^*}^m \right)
               \delta (\xi)
      = \sum_{n=0}^N \left( b_{n0} \partial_\xi^n
                     + (-)^n b_{n0}^* \partial_{\xi^*}^n \right)
                \delta (\xi) \, .
\end{equation}

\end{document}